\documentclass[runningheads]{llncs}
\usepackage{graphicx}
\usepackage[colorlinks=true,linkcolor=blue,citecolor=blue,urlcolor=blue]{hyperref}
\usepackage{multirow}
\usepackage{color}
\usepackage{amsmath}
\usepackage{marvosym}
\usepackage{hyperref}

\begin{document}
\title{FocalUNETR: A Focal Transformer for Boundary-aware Prostate Segmentation using CT Images}
\titlerunning{FocalUNETR}
%
\author{Chengyin Li\inst{1}\and 
Yao Qiang \inst{1}\and 
Rafi Ibn Sultan \inst{1} \and 
Hassan Bagher-Ebadian \inst{2} \and 
Prashant Khanduri  \inst{1}\and 
Indrin J. Chetty \inst{2} \and 
Dongxiao Zhu \inst{1}\textsuperscript{(\Letter)} 
}
\authorrunning{C. Li et al.}
%
\institute{Department of Computer Science, Wayne State University, Detroit MI, USA \\
\email{dzhu@wayne.edu}\\
\and
Department of Radiation Oncology, Henry Ford Cancer Institute, Detroit MI, USA}
\maketitle              
\begin{abstract}
Computed Tomography (CT) based precise prostate segmentation for treatment planning is challenging due to (1) the unclear boundary of the prostate derived from CT’s poor soft tissue contrast and (2) the limitation of convolutional neural network-based models in capturing long-range global context. Here we propose a novel focal transformer-based image segmentation architecture to effectively and efficiently extract local visual features and global context from CT images. Additionally, we design an auxiliary boundary-induced label regression task coupled with the main prostate segmentation task to address the unclear boundary issue in CT images. We demonstrate that this design significantly improves the quality of the CT-based prostate segmentation task over other competing methods, resulting in substantially improved performance, i.e., higher Dice Similarity Coefficient, lower Hausdorff Distance, and Average Symmetric Surface Distance, on both private and public CT image datasets. Our code is available at this \href{https://github.com/ChengyinLee/FocalUNETR.git}{link}.

\end{abstract}

\section{Introduction}
Prostate cancer is a leading cause of cancer-related deaths in adult males, as reported in studies, such as \cite{parikesit2016impact}. A common treatment option for prostate cancer is external beam radiation therapy (EBRT) \cite{d1998biochemical}, where CT scanning is a cost-effective tool for the treatment planning process compared with the more expensive magnetic resonance imaging (MRI). As a result, precise prostate segmentation in CT images becomes a crucial step, as it helps to ensure that the radiation doses are delivered effectively to the tumor tissues while minimizing harm to the surrounding healthy tissues.

Due to the relatively low spatial resolution and soft tissue contrast in CT images compared to MRI images, manual prostate segmentation in CT images can be time-consuming and may result in significant variations between operators \cite{li2023uncertainty}. Several automated segmentation methods have been proposed to alleviate these issues,  especially the fully convolutional networks (FCN) based U-Net \cite{ronneberger2015u} (an encoder-decoder architecture with skip connections to preserve details and extract local visual features) and its variants \cite{milletari2016v,xiao2018weighted,zhou2018unet++}.  Despite good progress, these methods often have limitations in capturing long-range relationships and global context information \cite{chen2021transunet} due to the inherent bias of convolutional operations. Researchers naturally turn to ViT \cite{Dosovitskiy2021AnII}, powered with self-attention (SA), for more possibilities: TransUNet first \cite{chen2021transunet} adapts ViT to medical image segmentation tasks by connecting several layers of the transformer module (multi-head SA) to the FCN-based encoder for better capturing the global context information from the high-level feature maps. TransFuse \cite{zhang2021transfuse} and MedT \cite{valanarasu2021medical} use a combined FCN and Transformer architecture with two branches to capture global dependency and low-level spatial details more effectively. Swin-UNet \cite{cao2021swin} is the first U-shaped network based purely on more efficient Swin Transformers \cite{liu2021swin} and outperforms models with FCN-based methods. UNETR \cite{hatamizadeh2022unetr} and SiwnUNETR \cite{tang2022self} are Transformer architectures extended for 3D inputs.

In spite of the improved performance for the aforementioned ViT-based networks, these methods utilize the standard or shifted-window-based SA, which is the fine-grained local SA and may overlook the local and global interactions \cite{yang2021focal,qiang2022attcat}. As reported by \cite{tang2022self}, even pre-trained with a massive amount of medical data using self-supervised learning, the performance of prostate segmentation task using high-resolution and better soft tissue contrast MRI images has not been completely satisfactory, not to mention the lower-quality CT images.  Additionally, the unclear boundary of the prostate in CT images derived from the low soft tissue contrast is not properly addressed \cite{he2021hf,wang2020boundary}.

\begin{figure*}[hbtp]
	\centering
	\includegraphics[width=\textwidth]{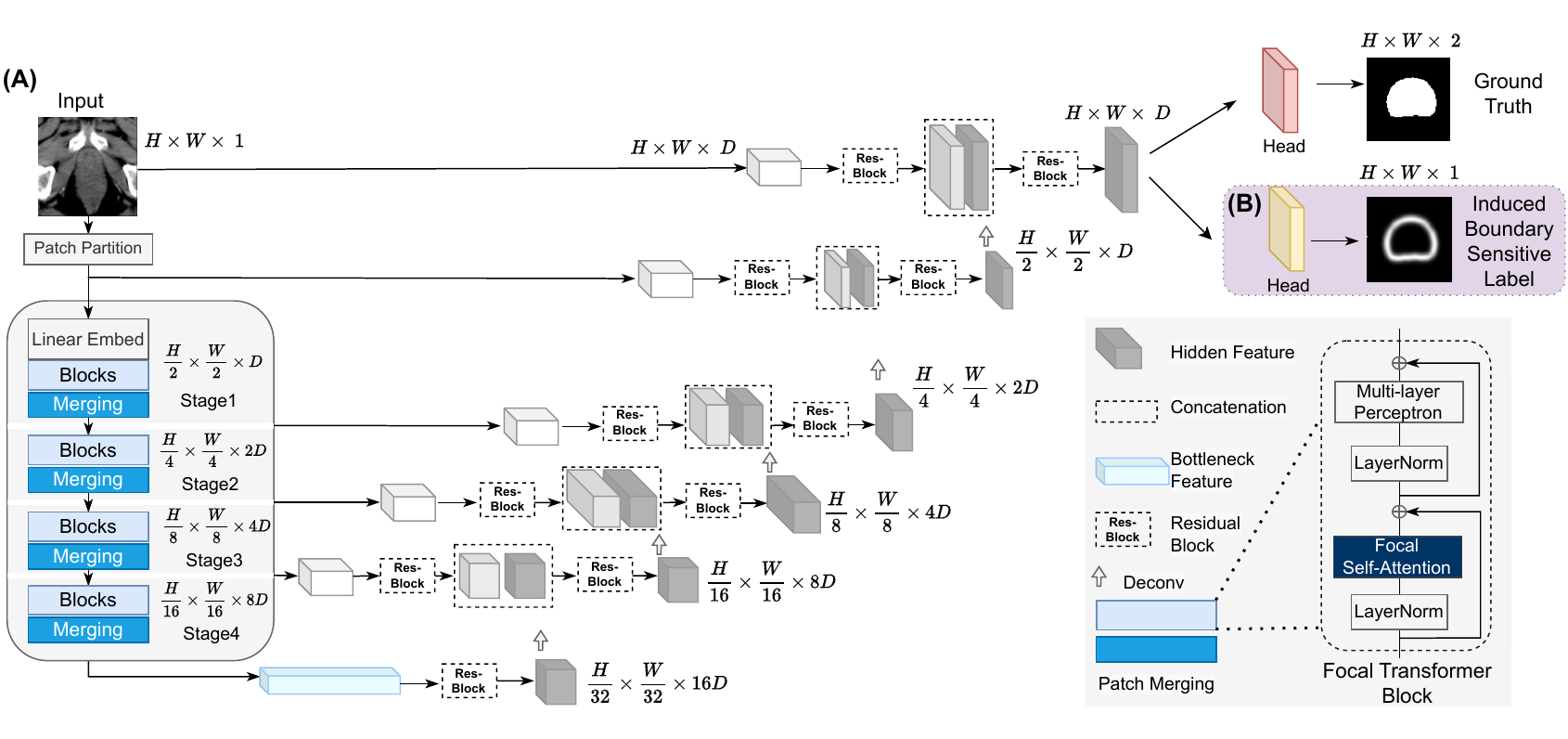}
	\caption{The architecture of FocalUNETR as (A) the main task for prostate segmentation and (B) a boundary-aware regression auxiliary task. 
	}\label{fig:focal_unetr}
\end{figure*}

Recently, Focal Transformer \cite{yang2021focal} is proposed for general computer vision tasks, in which focal self-attention is leveraged to incorporate both fine-grained local and coarse-grained global interactions. Each token attends its closest surrounding tokens with fine granularity, and the tokens far away with coarse granularity; thus, focal SA can capture both short- and long-range visual dependencies efficiently and effectively. Inspired by this work, we propose the FocalUNETR (Focal U-NEt TRansformers), a novel focal transformer architecture for CT-based medical image segmentation (Fig. \ref{fig:focal_unetr}A). Even though prior works such as Psi-Net \cite{murugesan2019psi} incorporates additional decoders to enhance boundary detection and distance map estimation, they either lack the capacity for effective global context capture through FCN-based techniques or overlook the significance of considering the randomness of the boundary, particularly in poor soft tissue contrast CT images for prostate segmentation. In contrast, our approach utilizes a multi-task learning strategy that leverages a Gaussian kernel over the boundary of the ground truth segmentation mask \cite{lin2021bsda} as an auxiliary boundary-aware contour regression task (Fig. \ref{fig:focal_unetr}B). This serves as a regularization term for the main task of generating the segmentation mask. And the auxiliary task enhances the model's generalizability by addressing the challenge of unclear boundaries in low-contrast CT images.

In this paper, we make several new contributions. First, we develop a novel focal transformer model (FocalUNETR) for CT-based prostate segmentation, which makes use of focal SA to hierarchically learn the feature maps accounting for both short- and long-range visual dependencies efficiently and effectively. Second, we also address the challenge of unclear boundaries specific to CT images by incorporating an auxiliary task of contour regression. Third, our methodology advances state-of-the-art performance via extensive experiments on both real-world and benchmark datasets.

\section{Methods}
\subsection{FocalUNETR}

\begin{figure*}[hbtp]
	\centering
        \scriptsize
        \includegraphics[scale=0.3]{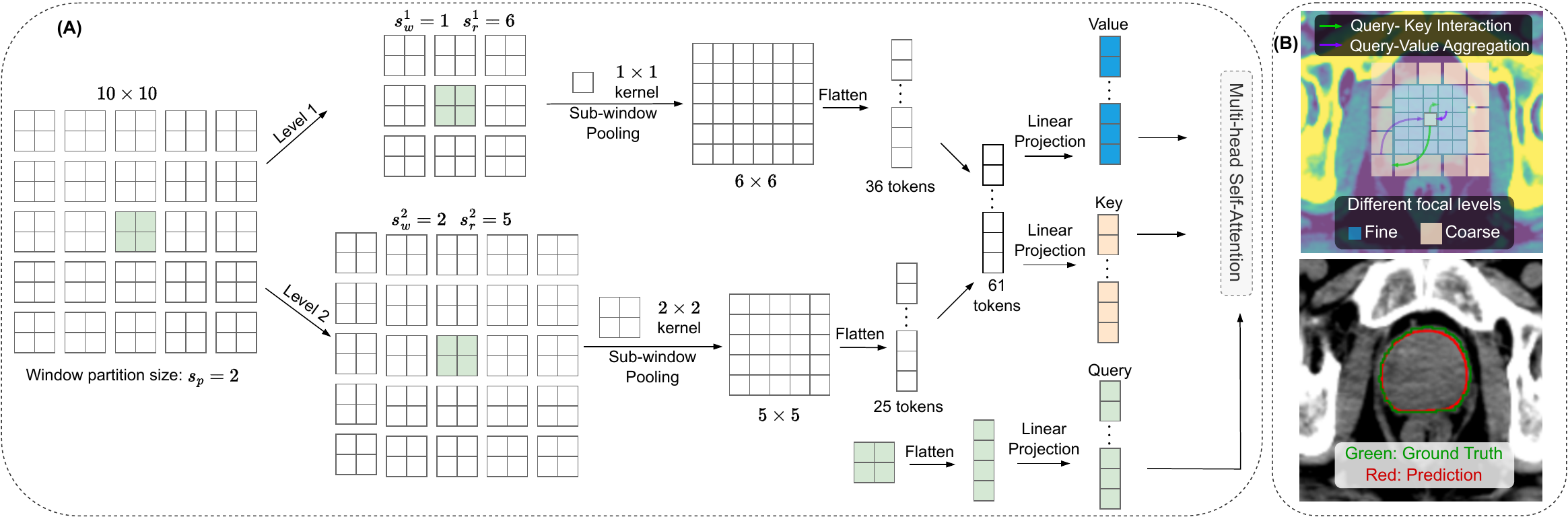}
	\caption{(A) The focal SA mechanism, and (B) an example of perfect boundary matching using focal SA for CT-based prostate segmentation task (lower panel), in which focal SA performs query-key interactions and query-value aggregations in both fine- and coarse-grained levels (upper panel).}
    \label{fig:focal_concept}
\end{figure*}

Our FocalUNETR architecture (Fig. \ref{fig:focal_unetr}) follows a multi-scale design similar to \cite{hatamizadeh2022unetr,tang2022self}, enabling us to obtain hierarchical feature maps at different stages. The input medical image $\mathcal{X} \in \mathcal{R}^{C \times H \times W}$ is first split into a sequence of tokens with dimension $\lceil \frac{H}{H'} \rceil \times \lceil \frac{W}{W'} \rceil$, where $H, W$ represent spatial height and width, respectively, and $C$ represents the number of channels. These tokens are then projected into an embedding space of dimension $D$ using a patch of resolution $(H', W')$. The SA is computed at two focal levels \cite{yang2021focal}: fine-grained and coarse-grained, as illustrated in Fig. \ref{fig:focal_concept}A. The focal SA attends to fine-grained tokens locally, while summarized tokens are attended to globally (reducing computational cost). We perform focal SA at the window level, where a feature map of $x \in \mathcal{R}^{d \times H''\times W''}$ with spatial size $H''\times W''$ and $d$ channels is partitioned into a grid of windows with size $s_w\times s_w$. For each window, we extract its surroundings using focal SA. 

For window-wise focal SA \cite{yang2021focal}, there are three terms $\{L, s_w, s_r\}$. Focal level $L$ is the number of granularity levels for which we extract the tokens for our focal SA. We present an example, depicted in Fig. \ref{fig:focal_concept}B, that illustrates the use of two focal levels (fine and coarse) for capturing the interaction of local and global context for optimal boundary-matching between the prediction and the ground truth for prostate segmentation. Focal window size $s_w^l$ is the size of the sub-window on which we get the summarized tokens at level $l \in \{1, \dots, L\}$. Focal region size $s_r^l$ is the number of sub-windows horizontally and vertically in attended regions at level $l$. The focal SA module proceeds in two main steps, sub-window pooling and attention computation. In the sub-window pooling step, an input feature map $x \in \mathcal{R}^{d \times H''\times W''}$ is split into a grid of sub-windows with size $\{s_w^l, s_w^l\}$, followed by a simple linear layer $f_p^l$ to pool the sub-windows spatially. The pooled feature maps at different levels $l$ provide rich information at both fine-grained and coarse-grained, where $x^l =  f_p^l(\hat{x}) \in \mathcal{R}^ {d \times \frac{H''}{s_w^l} \times \frac{W''}{s_w^l}}$, and $\hat{x} = \text{Reshape}(x) \in \mathcal{R}^{({d \times \frac{H''}{s_w^l} \times \frac{W''}{s_w^l}) \times (s_w^l \times s_w^l)}}$. After obtaining the pooled feature maps ${x^l}_1^L$, we calculate the query at the first level and key and value for all levels using three linear projection layers $f_q$, $f_k$, and $f_v$: 
$$ Q = f_q(x^1), K = \{K^l\}_1^L=f_k(\{x^1,\dots, x^L\}), V = \{V^l\}_1^L=f_v(\{x^1,\dots, x^L\}).$$ 

For the queries inside the $i$-th window $Q_i \in \mathcal{R}^{d \times s_w \times s_w}$, we extract the ${s_r^l \times s_r^l}$ keys and values from $K^l$ and $V^l$ around the window where the query lies in and then gather the keys and values from all $L$ to obtain $K_i = \{K_1, \dots, K_L\} \in \mathcal{R}^{s \times d}$ and $V_i = \{V_1, \dots, V_L\} \in \mathcal{R}^{s \times d}$, where $s$= $\sum_{l=1}^{L} (s_r^l)^2$. Finally, a relative position bias is added to compute the focal SA for $Q_i$ by $$\text{Attention}(Q_i, K_i, V_i) = \text{Softmax}(\frac{Q_i K_i^T}{\sqrt{d}}+B)V_i,$$  where $B = \{B^l\}_1^L$ is the learnable relative position bias \cite{yang2021focal}.

The encoder utilizes a patch size of $2 \times 2$ with a feature dimension of $2 \times 2 \times 1 = 4$ (i.e., a single input channel CT) and a $D$-dimensional embedding space. The overall architecture of the encoder comprises four stages of focal transformer blocks, with a patch merging layer applied between each stage to reduce the resolution by a factor of 2. We utilize an FCN-based decoder  (Fig. \ref{fig:focal_unetr}A) with skip connections to connect to the encoder at each resolution to construct a ``U-shaped" architecture for our CT-based prostate segmentation task. The output of the encoder is concatenated with processed input volume features and fed into a residual block. A final $1 \times 1$ convolutional layer with a suitable activation function, such as Softmax, is applied to obtain the required number of class-based probabilities.

\subsection{The Auxiliary Task}
For the main task of mask prediction (as illustrated in Fig. \ref{fig:focal_unetr}A), a combination of Dice loss and Cross-Entropy loss is employed to evaluate the concordance of the predicted mask and the ground truth on a pixel-wise level. The objective function for the segmentation head is given by:
$\mathcal{L}_{seg} = \mathcal{L}_{dice} (\hat{p}_i, G) + \mathcal{L}_{ce}(\hat{p}_i, G),$
where $\hat{p}_i$ represents the predicted probabilities from the main task and $G$ represents the ground truth mask, both given an input image $i$. The predicted probabilities, $\hat{p}_i$, are derived from the main task through the application of the FocalUNETR model to the input CT image. 


To address the challenge of unclear boundaries in CT-based prostate segmentation, an auxiliary task is introduced for the purpose of predicting boundary-aware contours to assist the main prostate segmentation task. This auxiliary task is achieved by attaching another convolution head after the extracted feature maps at the final stage (see Fig. \ref{fig:focal_unetr}B). The boundary-aware contour, or the induced boundary-sensitive label, is generated by considering pixels near the boundary of the prostate mask. To do this, the contour points and their surrounding pixels are formulated into a Gaussian distribution using a kernel with a fixed standard deviation of $\sigma$ (in this specific case, e.g., $\sigma = 1.6$) \cite{ma2020distance,he2021hf,lin2021bsda}. The resulting contour is a heatmap in the form of a $Heatsum$ function \cite{lin2021bsda}. We predict this heatmap with a regression task trained by minimizing mean-squared error instead of treating it as a single-pixel boundary segmentation problem. Given the ground truth of contour $G_i^C$, induced from the segmentation mask for input image $i$, and the reconstructed output probability $\hat{p}_i^C$, we use the following loss function:
$\mathcal{L}_{reg} = \frac{1}{N} \sum_i ||\hat{p}_i^C - G_i^C||_2$
where $N$ is the total number of images for each batch. This auxiliary task is trained concurrently with the main segmentation task.

A multi-task learning approach is adopted to regularize the main segmentation task through the auxiliary boundary prediction task. The overall loss function is a combination of $\mathcal{L}_{seg}$ and $\mathcal{L}_{reg}$:
$ \mathcal{L}_{tol} = \lambda_1 \mathcal{L}_{seg} + \lambda_2 \mathcal{L}_{reg}, $
where $\lambda_1$ and $\lambda_2$ are hyper-parameters that weigh the contribution of the mask prediction loss and contour regression loss, respectively, to the overall loss. The optimal setting of $\lambda_1 = \lambda_2 = 0.5$ is determined by trying different settings.

\section{Experiments and Results}
\subsection{Datasets and Implementation Details}
To evaluate our method, we use a large private dataset with 400 CT scans and a large public dataset with 300 CT scans (AMOS \cite{ji2022amos}). As far as we know, the AMOS dataset is the only publicly available CT dataset including prostate ground truth. We randomly split the private dataset with 280 scans for training, 40 for validation, and 80 for testing. The AMOS dataset has 200 scans for training and 100 for testing \cite{ji2022amos}. Although the AMOS dataset includes the prostate class, it mixes the prostate (in males) and the uterus (in females) into one single class labeled PRO/UTE. We filter out CT scans missing the PRO/UTE ground-truth segmentation.       

Regarding the architecture, we follow the hyperparameter settings suggested in \cite{yang2021focal}, with 2 focal levels, transformer blocks of depths [2, 2, 6, 2], and head numbers [4, 8, 16, 32] for each of the four stages. We then create FocalUNETR-S and FocalUNETR-B with $D$ as 48 and 64, respectively. These settings have 27.3 M and 48.3 M parameters, which are comparable to other state-of-the-art models in size. 

For the implementation, we utilize a server equipped with 8 Nvidia A100 GPUs, each with 40 GB of memory. All experiments are conducted in PyTorch, and each model is trained on a single GPU. We interpolate all CT scans into an isotropic voxel spacing of $[1.0 \times 1.0 \times 1.5]$ $mm$ for both datasets. Houndsfield unit (HU) range of $[-50, 150]$ is used and normalized to $[0, 1]$. Subsequently, each CT scan is cropped to a $128 \times 128 \times 64$ voxel patch around the prostate area, which is used as input for 3D models. For 2D models, we first slice each voxel patch in the axial direction into 64 slices of $128 \times 128$ images for training and stack them back for evaluation. For the private dataset, we train models for 200 epochs using the AdamW optimizer with an initial learning rate of $5e^{-4}$. An exponential learning rate scheduler with a warmup of 5 epochs is applied to the optimizer. The batch size is set to 24 for 2D models and 1 for 3D models. We use random flip, rotation, and intensity scaling as augmentation transforms with probabilities of 0.1, 0.1, and 0.2, respectively.  We also tried using 10\% percent of AMOS training set as validation data to find a better training parameter setting and re-trained the model with the full training set. However, we did not get improved performance compared with directly applying the training parameters learned from tuning the private dataset. We report the Dice Similarity Coefficient (DSC, \%), 95\% percentile Hausdorff Distance (HD, mm), and Average Symmetric Surface Distance (ASSD, mm) metrics.

\begin{table}[htbp]
\scriptsize
\centering
\caption{Quantitative performance comparison on the private and AMOS datasets with a mean (standard deviation) for 3 runs with different seeds. An asterisk (*) denotes the model is co-trained with the auxiliary contour regression task. The best results with/without the auxiliary task are boldfaced or italicized, respectively. }
\label{main_table1}
\resizebox{\textwidth}{!}{
\begin{tabular}{|l|ccc|ccc|}
\hline
\multirow{2}{*}{Method}  & \multicolumn{3}{c}{Private} & \multicolumn{3}{|c|}{AMOS} \\ \cline{2-7} 
 & DSC $\uparrow$ & HD $\downarrow$  & ASSD $\downarrow$ 
 & DSC $\uparrow$ & HD $\downarrow$ & ASSD $\downarrow$ \\ \cline{1-7}
 
 U-Net& 85.22 (1.23)  & 6.71 (1.03)  & 2.42 (0.65)      
 &  83.42 (2.28)  &   8.51 (1.56) & 2.79 (0.61)  \\
 
 UNet++ & 85.53 (1.61)  & 6.52 (1.13)  & 2.32 (0.58) 
 & 83.51 (2.31)    & 8.47 (1.62) & 2.81 (0.57)   \\
 
 AttUNet  & 85.61 (0.98)  & 6.57 (0.96) & 2.35 (0.72)  
 & 83.47 (2.34)   & 8.43  (1.85)   &   2.83 (0.59)  \\
 
 TransUNet & 85.75 (2.01)  & 6.43 (1.28) & 2.23 (0.67)  
 & 81.13 (3.03) & 9.32 (1.87)  & 3.71 (0.79)  \\
 
 Swin-UNet  & 86.25 (1.69)  & 6.29  (1.31)  & 2.15  (0.51) 
 & 83.35 (2.46)   & 8.61 (1.82)   & 3.20 (0.64)    \\ \hline

  U-Net (3D)  & 85.42 (1.34)  & 6.73 (0.93)  & 2.36 (0.67)      
 &  83.25 (2.37)  & 8.43 (1.65) & 2.86 (0.56) \\ 
 
 V-Net (3D)  & 84.42 (1.21)  & 6.65 (1.17) & 2.46 (0.61)
 & 81.02 (3.11) & 9.01 (1.93)  & 3.76 (0.82)   \\

 UNETR (3D)  & 82.21 (1.35)  & 7.25 (1.47) & 2.64 (0.75)
 & 81.09 (3.02) & 8.91 (1.86)  & 3.62 (0.79)   \\

 SwinUNETR (3D)  & 84.93 (1.26)  & 6.85 (1.21) & 2.48 (0.52)
 & 83.32 (2.23)  & 8.63 (1.62)  & 3.21 (0.68)   \\

 \hline
  nnUNet  & 85.86 (1.31)  & 6.43 (0.91) & 2.09 (0.53)
 & 83.56 (2.25) & 8.36 (1.77) & \textbf{\emph{2.65 (0.61)}} \\

\hline
FocalUNETR-S & 86.53 (1.65)  & 5.95 (1.29)  & 2.13 (0.29)  
 & 82.21 (2.67) &  8.73 (1.73) & 3.46 (0.75)   \\
 
 FocalUNETR-B & \emph{87.73 (1.36)}  & \emph{5.61 (1.18)}  & \emph{2.04 (0.23)}  
 & \emph{83.61 (2.18)} & \emph{8.32 (1.53)}  & 2.76 (0.69)  \\

\hline
 FocalUNETR-S*  & 87.84 (1.32)  & 5.59 (1.23)  & 2.12 (0.31)  
 &  83.24 (2.52) &  8.57 (1.70) &  3.04 (0.67)  \\
 
 FocalUNETR-B* & \textbf{89.23 (1.16)}  & \textbf{4.85 (1.05)}   & \textbf{1.81 (0.21)}  
 &   \textbf{83.79 (1.97)}  & \textbf{8.31 (1.45)} & 2.71 (0.62)  \\ \hline
\end{tabular}
}
\end{table}

\subsection{Experiments}
\subsubsection{Comparison with State-of-the-Art Methods.} To demonstrate the eﬀectiveness of FocalUNETR, we compare the CT-based prostate segmentation performance with three 2D U-Net-based methods: U-Net\cite{ronneberger2015u}, UNet++ \cite{zhou2018unet++}, and Attention U-Net (AttUNet) \cite{oktay2018attention}, two 2D transformer-based segmentation methods: TransUNet \cite{chen2021transunet} and Swin-UNet \cite{cao2021swin}, two 3D U-Net-based methods: U-Net (3D) \cite{cciccek20163d}  and V-Net \cite{milletari2016v}, and two 3D transformer-based models: UNETR \cite{hatamizadeh2022unetr} and SiwnUNETR \cite{tang2022self}. nnUNet \cite{isensee2019automated} is used for comparison as well. Both 2D and 3D models are included as there is no conclusive evidence for which type is better for this task \cite{wang2020boundary}. All methods (except nnUNet) follow the same settings as FocalUNETR and are trained from scratch. TransUNet and Swin-UNet are the only methods that are pre-trained on ImageNet. Detailed information regarding the number of parameters, FLOPs, and average inference time can be found in the supplementary materials.

Quantitative results are presented in Table \ref{main_table1}, which shows that the proposed FocalUNETR, even without co-training, outperforms other FCN and Transformer baselines (2D and 3D) in both datasets for most of the metrics.  The AMOS dataset mixes the prostate(males)/uterus(females, a relatively small portion). The morphology of the prostate and uterus is significantly different. Consequently, the models may struggle to provide accurate predictions for this specific portion of the uterus. Thus, the overall performance of FocalUNETR is overshadowed by this challenge, resulting in only moderate improvement over the baselines on the AMOS dataset. However, the performance margin significantly improves when using the real-world (private) dataset. When co-trained with the auxiliary contour regression task using the multi-task training strategy, the performance of FocalUNETRs is further improved.  In summary, these observations indicate that incorporating FocalUNETR and multi-task training with an auxiliary contour regression task can improve the challenging CT-based prostate segmentation performance.

\begin{figure*}[hbtp]
	\centering
	\includegraphics[scale=0.34]{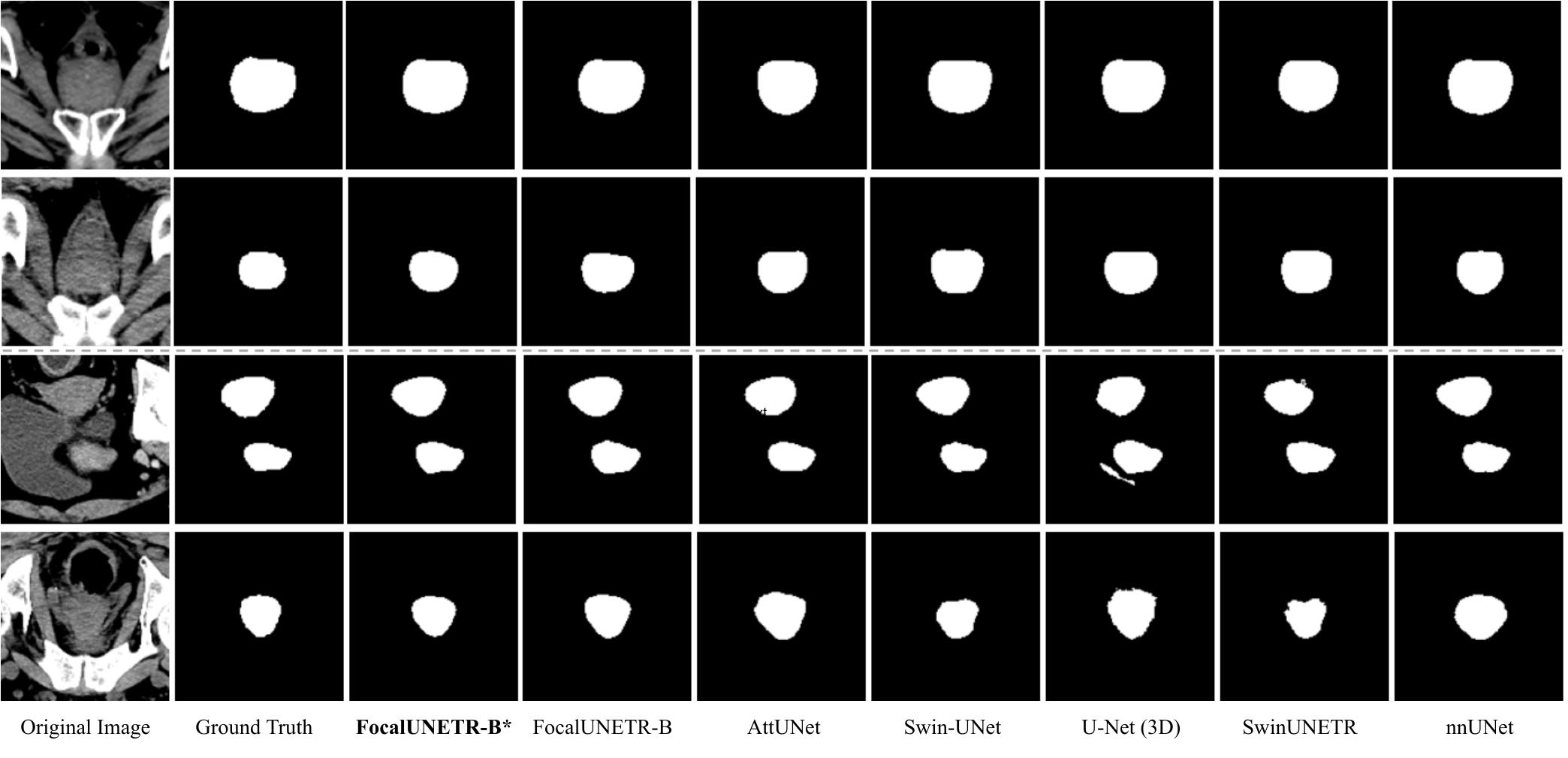}
	\caption{Qualitative results on sample test CT images from the private (first two rows) and AMOS (last two rows) datasets}
 \label{fig:quality_r_1}
\end{figure*}

Qualitative results of several representative methods are visualized in Fig. \ref{fig:quality_r_1}. The figure shows that our FocalUNETR-B and FocalUNETR-B* generate more accurate segmentation results that are more consistent with the ground truth than the results of the baseline models. All methods perform well for relatively easy cases ($1^{st}$ row in Fig. \ref{fig:quality_r_1}), but the FocalUNETRs outperform the other methods. For more challenging cases (rows 2-4 in Fig. \ref{fig:quality_r_1}), such as unclear boundaries and mixed PRO/UTE labels, FocalUNETRs still perform better than other methods. Additionally, the FocalUNETRs are less likely to produce false positives (see more in supplementary materials) for CT images without a foreground ground truth, due to the focal SA mechanism that enables the model to capture global context and helps to identify the correct boundary and shape of the prostate. Overall, the FocalUNETRs demonstrate improved segmentation capabilities while preserving shapes more precisely, making them promising tools for clinical applications.

\begin{table}[hbtp]
\centering
\caption{Ablation study on different settings of total loss for FocalUNETR-B on the private dataset}
\label{main_table2}
\begin{tabular}{|l|c|c|c|c|}
\hline
$\mathcal{L}_{tol}$  & $\mathcal{L}_{seg}$ & $0.8\mathcal{L}_{seg}+0.2\mathcal{L}_{reg}$ & $0.5\mathcal{L}_{seg}+0.5\mathcal{L}_{reg}$ & $0.2\mathcal{L}_{seg}+0.8\mathcal{L}_{reg}$ \\
\hline
DSC $\uparrow$     & 87.73 $\pm$ 1.36  & 88.01  $\pm$ 1.38     & \textbf{89.23 $\pm$ 1.16}    & 87.53  $\pm$ 2.13\\
\hline
\end{tabular}
\end{table}

\subsubsection{Ablation Study.} To better examine the efficacy of the auxiliary task for FocalUNETR, we selected different settings of $\lambda_1$ and $\lambda_2$ for the overall loss function $\mathcal{L}_{tol}$ on the private dataset. The results (Table \ref{main_table2}) indicate that as the value of $\lambda_2$ is gradually increased and that of $\lambda_1$ is correspondingly decreased (thereby increasing the relative importance of the auxiliary contour regression task), segmentation performance initially improves. However, as the ratio of contour information to segmentation mask information becomes too unbalanced, performance begins to decline. Thus, it can be inferred that the optimal setting for these parameters is when both $\lambda_1$ and $\lambda_2$ are set to 0.5.

\section{Conclusion}

In summary, the proposed FocalUNETR architecture has demonstrated the ability to effectively capture local visual features and global contexts in CT images by utilizing the focal self-attention mechanism. The auxiliary contour regression task has also been shown to improve the segmentation performance for unclear boundary issues in low-contrast CT images. Extensive experiments on two large CT datasets have shown that the FocalUNETR outperforms state-of-the-art methods for the prostate segmentation task. Future work includes the evaluation of other organs and extending the focal self-attention mechanism for 3D inputs. 


%
%
%
\newpage
\bibliographystyle{splncs04}
\bibliography{refs}

\newpage
\clearpage
\setcounter{table}{0}
\setcounter{figure}{0}
\section{Appendix}
\begin{table}
\centering
\caption{The number parameters, FLOPs, and average inference time per case for different models: our FocalUNETR shows a comparable model size, relatively small FLOPs, and fast inference speed to most of the SOTAs.}\label{tab1}
\scriptsize
\begin{tabular}{|l|c|c|c|}
\hline
Model & Param. (M) & FLOPs (G) & Average Inference Time (s)\\
\hline
U-Net & 7.2 & 9.3 & 3.12\\
UNet ++ & 22.5 & 60.4 & 4.31\\
AttUNet & 19.8 & 25.5 & 3.53\\
TransUNet & 105.3 & 29.3 & 4.87\\
Swin-UNet & 41.4 & 9.0 & 3.58\\
U-Net (3D) & 16.6 & 285 & 6.51\\
V-Net (3D) & 45.6 & 586 & 6.72\\
UNETR (3D) & 92.6 & 75.4 & 6.49\\
SwinUNETR (3D) & 62.2 & 350 & 7.23\\
nnUNet & 19.3& 389 & 9.65 \\
\hline
FocalUNETR-S & 27.3 & 15.7 & 4.36\\
FocalUNETR-B & 48.3 & 27.5 & 5.35\\
\hline
\end{tabular}
\end{table}

\begin{figure}
\centering
\includegraphics[scale=0.5]{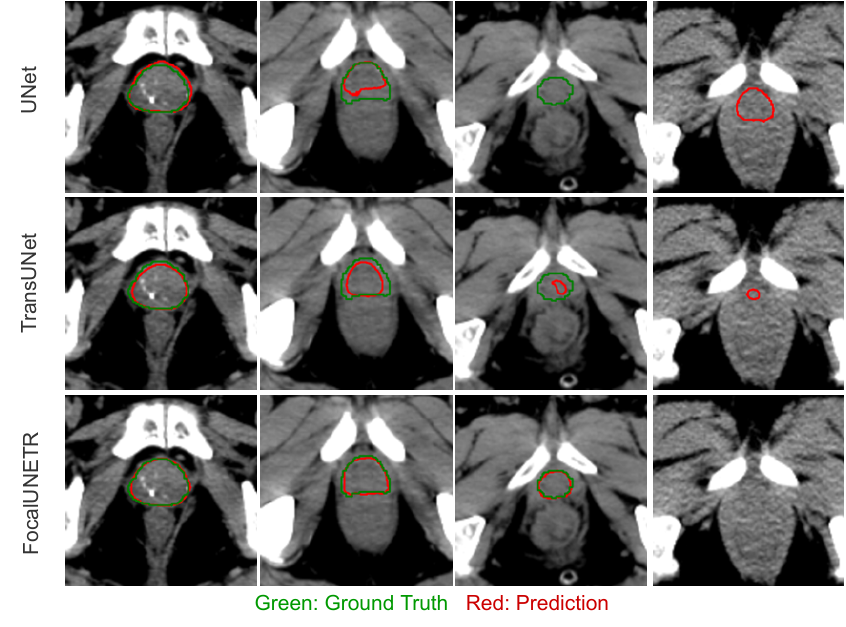}
\caption{Qualitative results of prostate segmentation by comparing our FocalUNETR-B with UNet and TransUNet in 2D settings. All methods perform well for easy cases, but our FocalUNETR-B can be even better. FocalUNETR-B is less likely to give a false prediction (false positives) for CT images without a foreground mask.} \label{fig1}
\end{figure}

\end{document}